\documentclass[12pt,american]{article}
\usepackage[T1]{fontenc}
\usepackage[latin1]{inputenc}
\usepackage{geometry}
\usepackage{amsmath}
\usepackage{babel}

\makeatletter

\providecommand{\LyX}{L\kern-.1667em\lower.25em\hbox{Y}\kern-.125emX\@}

\newcommand{\lyxaddress}[1]{
  \par {\raggedright #1 
  \vspace{1.4em}
  \noindent\par}
}

\usepackage{axodraw}
\makeatother
\makeatother

\begin{document}

\title{\hfill{}\\
A rolling tachyon field for both dark energy and dark halos of galaxies}

\author{M.B.Causse\thanks{
causse@gamum2.in2p3.fr
} }

\maketitle

\lyxaddress{Groupe d'Astroparticules de Montpellier, IN2P3/CNRS, Université Montpellier
II Place E.Bataillon, CC085, 34095 Montpellier Cedex 5 France}

\begin{abstract}
Distance measurements to type Ia supernovae (SNe Ia) indicate that the universe
is accelerating and it is spatially flat. Otherwise, the rotation curves for
galaxies or galaxy clusters reveal the existence of a dark matter component.
Then, approximately 0.35 of the total energy density of the universe consists
of dark matter, and 0.65 of the energy density corresponds to dark energy. These
two dark components of the universe have different properties such that: dark
matter clusters gravitationally at galactic scales while dark energy, dominating
at large scale, does not cluster gravitationally. It seems difficult to find
a candidate which explains the observations at both large and small scales.
In this paper we take into account a scalar field, more precisely a rolling
tachyon arising from string theory, which seems be capable to explain the observations
at different scales. We examine this scalar field behavior at galactic scales
and we fit the rotational curve data of spiral galaxies. Our results are comparable
with the curves obtained on one hand from a complex scalar field, which only
plays the role of the dark matter at galactic scales, and on the other hand
from the MOND model.
\end{abstract}

\section{Introduction}

The observational data of our Universe, at different scales, reveal that our
Universe is dominated by a Dark density component. Indeed, the measurement of
type Ia supernova light curves by two groups, the Supernova Cosmology Project
\cite{1} and the High-z Supernova Team \cite{2}, leads to the conclusion:
the Universe is speeding up, not slowing down. Otherwise, an independent line
of evidence for the accelerating Universe comes from measurements of the composition
of the universe. The Cosmic Microwave Background (CMB) anisotropy measurements
indicate that the Universe is flat \cite{3}, so the total density is: \( \Omega _{c}=1. \)
Moreover, in a flat universe the matter density (\( \Omega _{M} \)) and energy
density (\( \Omega _{DE} \)) must sum to the critical density: \( \Omega _{M}+\Omega _{DE}\: =\Omega _{c} \),
where matter contribution is: \( \Omega _{M}=0.33\pm 0.04 \) \cite{4}, leaving
two thirds of the dark energy (\( \Omega _{DE}\simeq 0.65 \)). In order to
have escaped detection, this dark energy must be smoothly distributed. In order
not to interfere with the formation of cosmic structure, the energy density
in this component must change more slowly than matter, so that it was subdominant
in the past. Theorists have been very busy suggesting all kinds of interesting
possibilities for the dark energy. Dark energy is a negative pressure component
with an equation of state as \( p=\omega \, \rho  \) with \( \omega  \) negative,
not necessary constant. One of the possible explanations is the existence of
a non-zero vacuum energy, i.e a ``cosmological constant'' where \( \omega =-1 \).
For candidates like homogeneous tachyon field(\cite{5}) or scalar field (quintessence)
\cite{6}, \( \omega  \) is time dependent and can vary between \( -1 \) and
\( +0 \).

At small scales, Astronomers studying the motions of stars in spiral galaxies
noticed that the mean star velocity did not drop off with radius from the galactic
center as rapidly as the fall-off in luminous mass in the galaxy dictated according
to Newtonian gravity. The stars far from the center were rotating too fast to
be balanced by the gravitational force from the luminous mass contained within
that radius. This led to the proposition that most of the mass in a galaxy was
low luminosity mass of some kind, and this invisible mass was called Dark Matter
(DM)\cite{7}.The amount of dark matter present in the universe has been estimated
using various techniques, including observing the velocities of galaxies in
clusters \cite{8}and calculating the gravitational mass of galactic clusters
by their gravitational lensing effects on surrounding spacetime. Then, the matter
density is the sum of baryonic density \( \Omega _{B}=0.05 \) and dark matter
density \( \Omega _{DM}=0.25 \). So, the rotation curves (RCs) of spiral galaxies
\cite{9} are the most natural way to model dark matter. Contrary to the dark
energy component, the dark matter component , with an equation of state: \( p=0 \),
clusters gravitationally at galactic scales. Particle physics provides an attractive
solution to the non baryonic dark matter problem. Long-lived or stable particles
with very weak interactions can remain from the earliest moments of the universe
in sufficient numbers to account for a significant fraction of critical density. 

Due to the supersymmetry (SUSY) breaking mechanisms, the supersymmetric theories
offer the most promising explanations concerning the dark energy and cold dark
matter components of the universe. However, the nature of the supersymmetric
candidates depend on the scale. Indeed, in the Minimal Supersymmetric Standard
Model (MSSM), where soft SUSY breaking terms are present, the lightest neutralino
is a good candidate for the cold dark matter \cite{10} but not for the dark
energy. The dark energy is due to a scalar field, the quintessence field. And,
the determination of the shape of the quintessence potential depends on the
SUSY breaking\footnote{%
However, SUSY breaking scale, at least of the order of 1 Tev, is the crucial
point to acheive in explicit string models. Indeed, such a breaking must arise
from nonperturbative effects which are often difficult to control \cite{11}. 
} mechanism. Then, the quintessence field is not a good candidate to explain
the observations at both small and large scales (although, some scalar fields
can play the dark matter role \cite{17} only ). Recent work on tachyon field
(arising from string theory) in cosmology (\cite{5},\cite{12},\cite{13} and
\cite{14}) tends to show that a tachyon field can behave like dark energy with
negative pressure at large scale and can cluster gravitationally on small scales.
So, if the fine-tuning problem can be explained,the tachyon field is a viable
candidate to explain the observations at both small and large scales. In this
context, we consider a tachyon field model, given by T. Padmanabhan and T.R.Choudhury
in reference \cite{14}, thanks to which it is possible to obtain different
equations of state at different scales. At large scales, the authors have shown
that this model reproduces the correct behavior. Our study concerns the behavior
of this field on small scales and more precisely its contribution to the dark
matter component of the universe. By simplicity we retain only the Newtonian
solution and fit the rotation curve data of spiral galaxies. 

This paper is organized as follows. In section 2, we quickly review the tachyon
scalar field model and develop the Newtonian solution of the tachyonic Lagrangian
considered. In section 3, we present the plots of the spiral galaxies rotation
curves obtained with this type of tachyonic dark matter. Finally, in section
4 we conclude.

\section{The tachyon scalar field model}

In the string theoretical context, a tachyonic Lagrangian arises naturally.This
is a generalization of the Lagrangian for a relativistic particle and it takes
the following form:

\begin{equation}
\label{equ1}
L_{tach}=-V(\phi )\sqrt{1-\partial ^{i}\phi \partial _{i}\phi }
\end{equation}

where \( \phi  \) is the tachyon field and \( V(\phi ) \) the potential.

This theory admits solutions where the two following conditions: \( V\, \rightarrow \, 0,\: \partial _{i}\phi \partial ^{i}\phi \, \rightarrow \, 1 \)
can be met simultaneously. Such solutions have finite momentum density and energy
density. Then, the solutions can depend on both space and time and the momentum
density can be an arbitrary function of the spatial coordinate. Thus, this context
seems to be adapted at the same time to large and small scales in cosmology.

\subsection{The model}

The model considered is given by T. Padmanabhan and T.R.Choudhury in reference
\cite{14}. We quickly review the principal ingredients. For more details, the
reader is referred to \cite{14}. This model takes into account the gravitational
interaction modified by a scalar field and a scalar potential. The effective
low energy action is:

\begin{equation}
\label{equ2}
S=\int d^{4}x\sqrt{-g}\, (\frac{R}{16\pi G}\, -V(\phi )\sqrt{1-\partial ^{i}\phi \partial _{i}\phi }\, )
\end{equation}

where \( R \) is the curvature scalar, \( G \) the gravitation constant (\( \overline{h}=c=1 \))
and \( g \) the determinant of the metric. Moreover, the field \( \phi  \)
is regarded as a scalar field for simplicity. The Einstein equations are

\begin{equation}
\label{equ3}
R^{i}_{k}\, -\frac{1}{2}\delta ^{i}_{k}R\, =8\pi GT_{k}^{i}
\end{equation}

and the stress tensor for the scalar field can be written in a perfect fluid
form

\begin{equation}
\label{equ4}
T^{i}_{k}\, =(\rho +p)u^{i}u_{k}-p\delta ^{i}_{k}\; \; \; with\; u_{k}=\frac{\partial _{k}\phi }{\sqrt{\partial ^{i}\phi \partial _{i}\phi }}\; \; and\; \; u_{k}u^{k}=1
\end{equation}

The density and the pressure are respectively:

\begin{equation}
\label{equ5}
\rho =\frac{V(\phi )}{\sqrt{1-\partial ^{i}}\phi \partial _{i}\phi }\; ;\; \; p=-V(\phi )\sqrt{1-\partial ^{i}\phi \partial _{i}\phi }
\end{equation}

This stress tensor can be put in the form according to:

\begin{equation}
\label{equ6}
\rho =\rho _{DE}+\rho _{DM}
\end{equation}

\[
p=p_{DE}+p_{DM}\]

with the two different equations of state:
\begin{equation}
\label{equ7}
\rho _{DM}=\frac{V(\phi )\partial ^{i}\phi \partial _{i}\phi }{\sqrt{1-\partial ^{i}}\phi \partial _{i}\phi }\; and\; p_{DM}=0
\end{equation}

\[
\rho _{DE}=V(\phi )\sqrt{1-\partial ^{i}\phi \partial _{i}\phi }\; and\; p_{DE}=-\rho _{DE}\]

Thus, we notice that the stress tensor can be broken up into the sum of two
components, one having the behavior of dark energy and the other one the dark
matter behavior. However, the scalar field \( \phi  \) must have a particular
configuration to account for the equations of state on different scales.Then,
the effective field\footnote{%
The effective field is the average of a field \( \phi (t,x) \) over a length
scale \( r \) (\cite{12}, \cite{13} and \cite{16})
} that we chose has the following form:

\begin{equation}
\label{equ8}
\overline{\phi }(t,\, r)=A(r)t+f(r)\exp (-2t)
\end{equation}

where the functions, \( A(r) \) and \( f(r) \), determine the evolution of
the field on the various scales. The time dependence is related to the potential

\begin{equation}
\label{equ9}
\overline{V_{r}}(\overline{\phi }(t,\, r))=V_{0}\exp (-\frac{\overline{\phi }(t,\, r)}{\phi _{0}})
\end{equation}

At large scales, the rate of the expansion of the universe is determined by
\( A(r) \). Indeed, when \( r \) increases the fluctuations decrease, so \( \phi (r) \)
will be a decreasing function of \( r \) thus \( A(r) \) will have a value
less than unity. Taking \( \phi ^{\cdot }(r)=A(r)=constant\; \, and\, \; V=V_{0}/\exp (A(r)t) \)
one can find consistent set of solutions for an \( \Omega =1 \) Friedmann-Robertson-Walker
model with a power expansion \( a(t)\propto t^{n} \) and \( A(r)=\sqrt{\frac{2}{3n}} \).

For the small scales (\( r \) small), the dust component prevails. So we consider
the asymptotic limit (\( t\rightarrow \infty  \)) and \( \rho _{DE}\approx 0 \)
what implies that: \( V\rightarrow 0\; and\; \partial _{i}\phi \partial ^{i}\phi \rightarrow 1,\; \, thus\, \; A(r)\rightarrow 1 \).
In this context we have

\begin{equation}
\label{equ10}
\rho _{DM}\approx \frac{V_{0}}{\sqrt{4f(r)}}\; for\; \phi _{0}\sim 1/2\; and\, \; \rho _{DE}\approx 0
\end{equation}

the pressureless component dominates at galactic scales and the associated density
is independent of time (\cite{12}, \cite{13}). This resembles the non-interacting
dark matter and the field \( \overline{\phi }(t,\, r) \) can play the role
of dark matter. 

We have just seen that this scalar field can give an account at the same time
of dark energy and dark matter components of the universe. The next step is
now to study the behavior of this field at galactic scales.

\subsection{Newtonian solution}

In this section, we consider the scalar field \( \phi  \) as the dark matter
of the universe at galactic scales. Under these conditions, the expression of
the field takes the following form:

\begin{equation}
\label{equ11}
\overline{\phi }(t,\, r)=A(r)t+f(r)\exp (-2t)\; \, with\, \; A(r)\rightarrow 1
\end{equation}

The analytical expression of \( f(r) \) is given by the solution of the differential
equation for the scalar field. This equation is obtained from the Lagrangian
(\ref{equ2}), the Einstein equations (\ref{equ3}) and (\ref{equ4}) and the
Klein Gordon equation for a massless field. Moreover, the spherically symmetric
solutions are characterized by the Schwarzschild metric:
\begin{equation}
\label{eq12}
ds^{2}=\exp (\nu (r))dt^{2}-\exp (\lambda (r))dr^{2}-r^{2}(d\theta ^{2}+\sin ^{2}\theta d\varphi ^{2})
\end{equation}

where \( \nu (r)\: \: and\: \: \lambda (r) \) are constant for the Newtonian
solutions.

Then, the differential equation for the scalar field is given by

\begin{equation}
\label{equ13}
f^{''}(r)+\frac{2}{r}f^{'}(r)-4f(r)=0
\end{equation}
 where \( '=\frac{d}{dr} \). For simplicity, we take \( \exp (\lambda (r)-\nu (r))=1 \)
since \( \nu (r)\: \: and\: \: \lambda (r) \) are constant.

The Newtonian form of \( f(r) \) reads 

\begin{equation}
\label{equ14}
f(r)=\frac{1}{r}(C_{1}\sinh (2r)+C_{2}\cosh (2r))
\end{equation}

where \( C_{1}\: \: and\: \: C_{2} \) are two constants, with \( C_{2}=0 \),
so that the solution is not singular at the origin. Taking into account the
equations (\ref{equ14}) and (\ref{equ10}) the density energy expression takes
the following form:

\begin{equation}
\label{equ15}
\rho _{DM}=\frac{V_{0}}{\sqrt{4f(r)}}=\frac{V_{0}}{2\sqrt{C_{1}}}\sqrt{\frac{r}{\sinh (2r)}}
\end{equation}

For small \( r \), we have \( \rho \propto \frac{V_{0}}{2\sqrt{2C_{1}}}(1-\frac{r^{2}}{3}+\frac{r^{4}}{10}) \),
i.e. it is proportional to a constant \( C=\frac{V_{0}}{2\sqrt{2C_{1}}} \).
This constant resembles a core radius. Otherwise, the mass function is given
by \( M(x)=\int _{0}^{x}\rho (r)r^{2}dr \) then we obtain

\begin{equation}
\label{16}
M(x)=C\int ^{x}_{0}\frac{r^{(5/2)}}{\sqrt{\sinh }(2r)}
\end{equation}

Therefore, we shall model rotation curves of spiral galaxies. Observational
data show that rotation curves are becoming flat in the surrounding region of
galaxies where data are received from the 21cm wavelength of neutral hydrogen
(HI). We introduce dark matter consisting of the massless scalar field which
interacts with the luminous matter exclusively by the gravitational force. We
apply only Newtonian solutions of our model. 

For the static spherically symmetric metric (\ref{eq12}) considered here, circular
orbit geodesics obey to the velocity \( v_{\phi }^{2}\simeq \frac{M(x)}{x} \)
then:

\begin{equation}
\label{17}
v^{2}_{\phi }\simeq \frac{C}{x}\int ^{x}_{0}\frac{r^{(5/2)}}{\sqrt{\sinh }(2r)}
\end{equation}

The behavior of \( v_{\phi } \) is presented in figure 1, for \( C'=90km/s \)
(with \( C'=\sqrt{C} \) ). At low \( x \), this curve presents a maximum and
an asymptotic value for large \( x \).

\begin{center}
\begin{figure}[htbp]
\vspace{0.01cm}
\hspace{-0.05cm}
\epsfxsize=\dimen116
\epsfysize=\dimen116
\epsfbox[98 459 506 708]{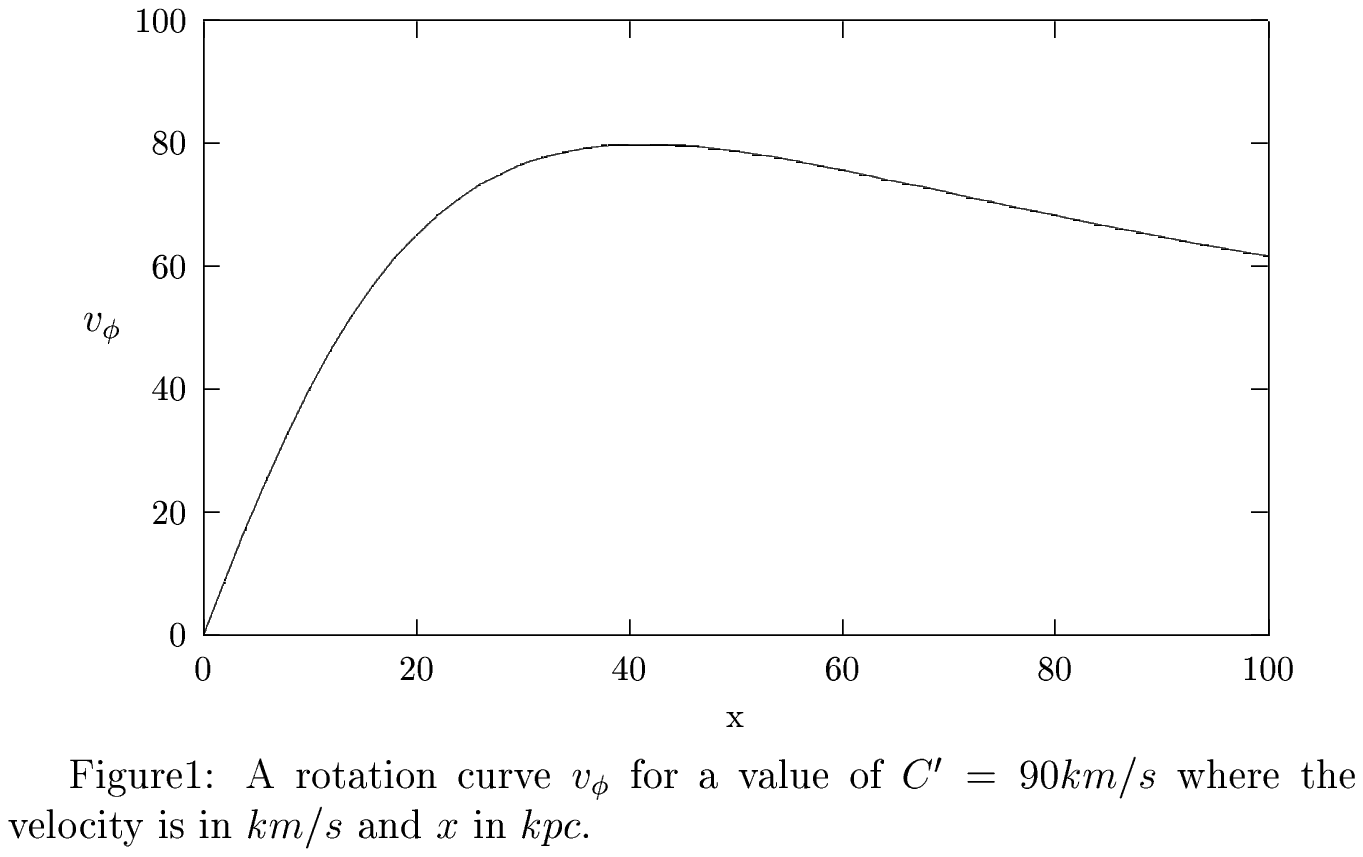}

\end{figure}
\end{center}

\section{Rotation curves for spiral galaxies}

To model rotational curves of spiral galaxies, we use the universal rotation
curves of Persic, Salucci and Stel \cite{9}\footnote{%
However, universal rotation curves, depending only on the luminosity, don't
represent very well the data in some cases\cite{18}
} and some individual ones \cite{15}. The total rotation curves result from
a combination of several components, i.e. the stellar disk, the halo, the gas
and the bulge of the galaxy

\begin{equation}
\label{18}
v_{total}=\sqrt{v^{2}_{disk}+v^{2}_{halo}+v^{2}_{gas}+v^{2}_{bulge}}
\end{equation}
 The rotation curve for stellar disk follows from an exponential thin disk light
distribution and the halo contribution is given by \( v_{\phi } \).

\subsection{Universal rotation curves}

The universal rotation curves \cite{9} depend on two components: the stellar
disk and our halo

\begin{equation}
\label{19}
v_{total}=\sqrt{v^{2}_{disk}+v^{2}_{halo}}
\end{equation}

The contribution from the stellar disk can be written as:

\begin{equation}
\label{20}
v^{2}_{disk}(x)=V^{2}(R_{opt})\beta \frac{1.97\, x^{(1.22)}}{(\, x^{2}+0.78^{2}\, )^{1.43}}
\end{equation}

where \( x=R/R_{opt} \)\footnote{%
\( R_{opt} \) is the optical radius such as \( R_{opt}\equiv 3.2R_{D} \) where
\( R_{D} \) is the disk exponential length-scale.
} , \( v^{2}_{halo}=v^{2}_{\phi } \) and \( \beta =0.72+\frac{0.35}{2.5}(M_{B}^{*}-M_{B}) \).
The absolute magnitude in the blue band is \( M_{B}^{*}=-20.5 \) and \( M_{B}=-0.38+0.92M_{I} \)
(\( M_{I} \) for the \( I \) band). The constants in the \( \beta  \) function
are arranged in order to obtain the best fit for our rotational curves. The
equation (\ref{20}) is valid within the range \( 0.04\simeq x\leq 2 \). 

\begin{center}
\begin{figure}[htbp]
\vspace{0.05cm}
\hspace{-0.25cm}
\epsfxsize=\dimen116
\epsfysize=\dimen116
\epsfbox[84 220 540 790]{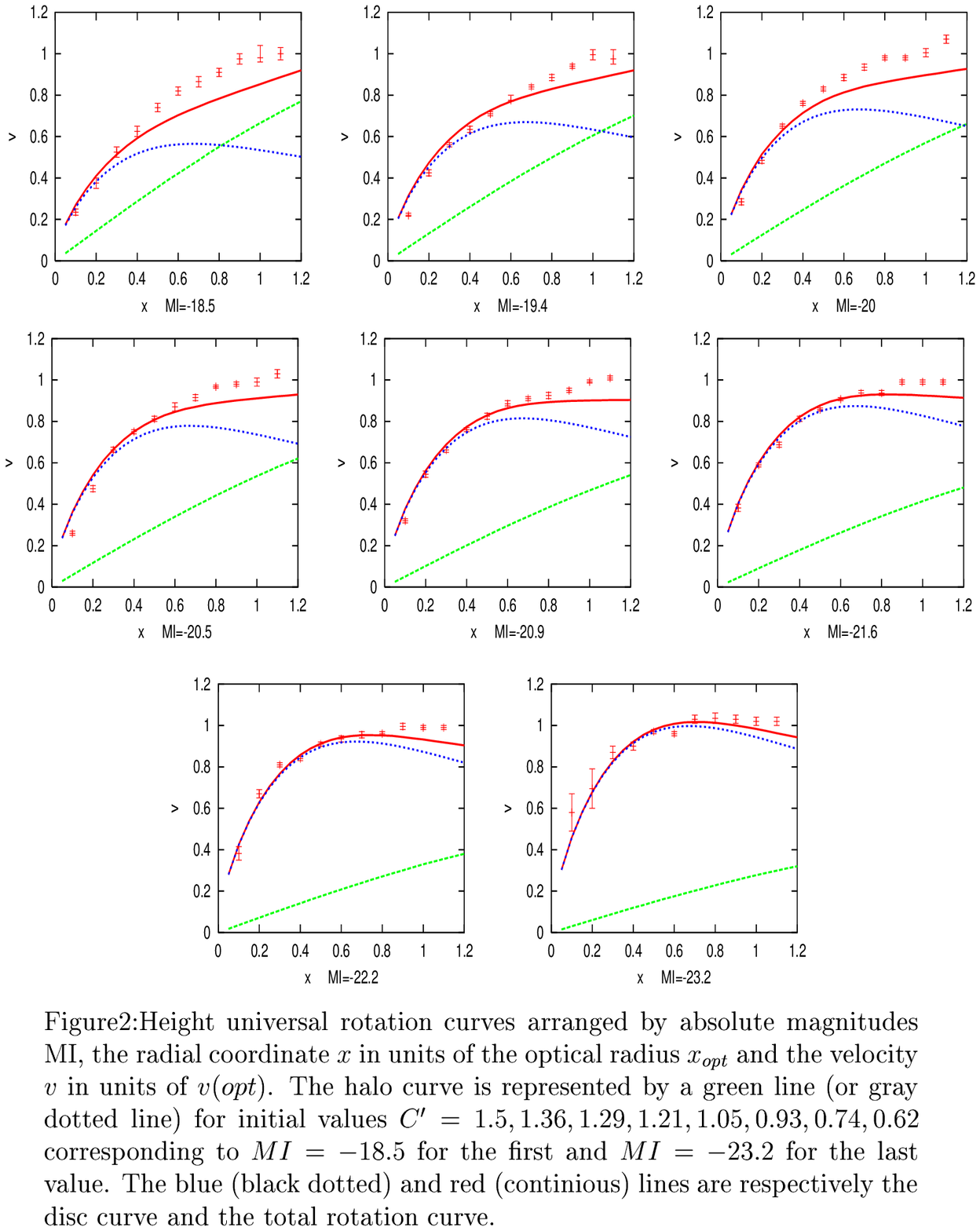}

\end{figure}
\end{center}

\begin{center}
\begin{figure}[htbp]
\vspace{0.05cm}
\hspace{-0.25cm}
\epsfxsize=\dimen116
\epsfysize=\dimen116
\epsfbox[75 225 540 769]{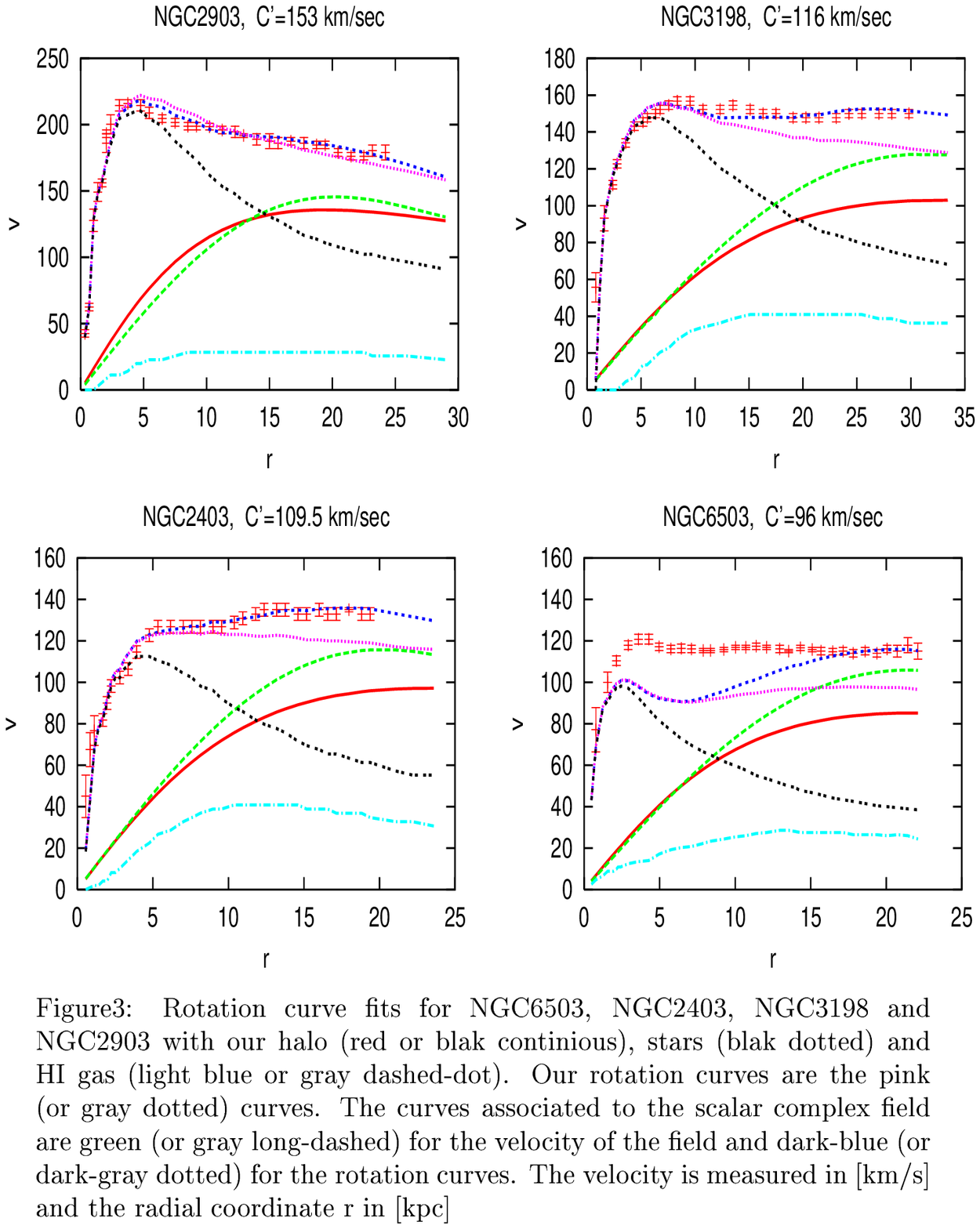}

\end{figure}
\end{center} 

\begin{center}
\begin{figure}[htbp]
\vspace{0.05cm}
\hspace{-0.25cm}
\epsfxsize=\dimen116
\epsfysize=\dimen116
\epsfbox[78 501 540 842]{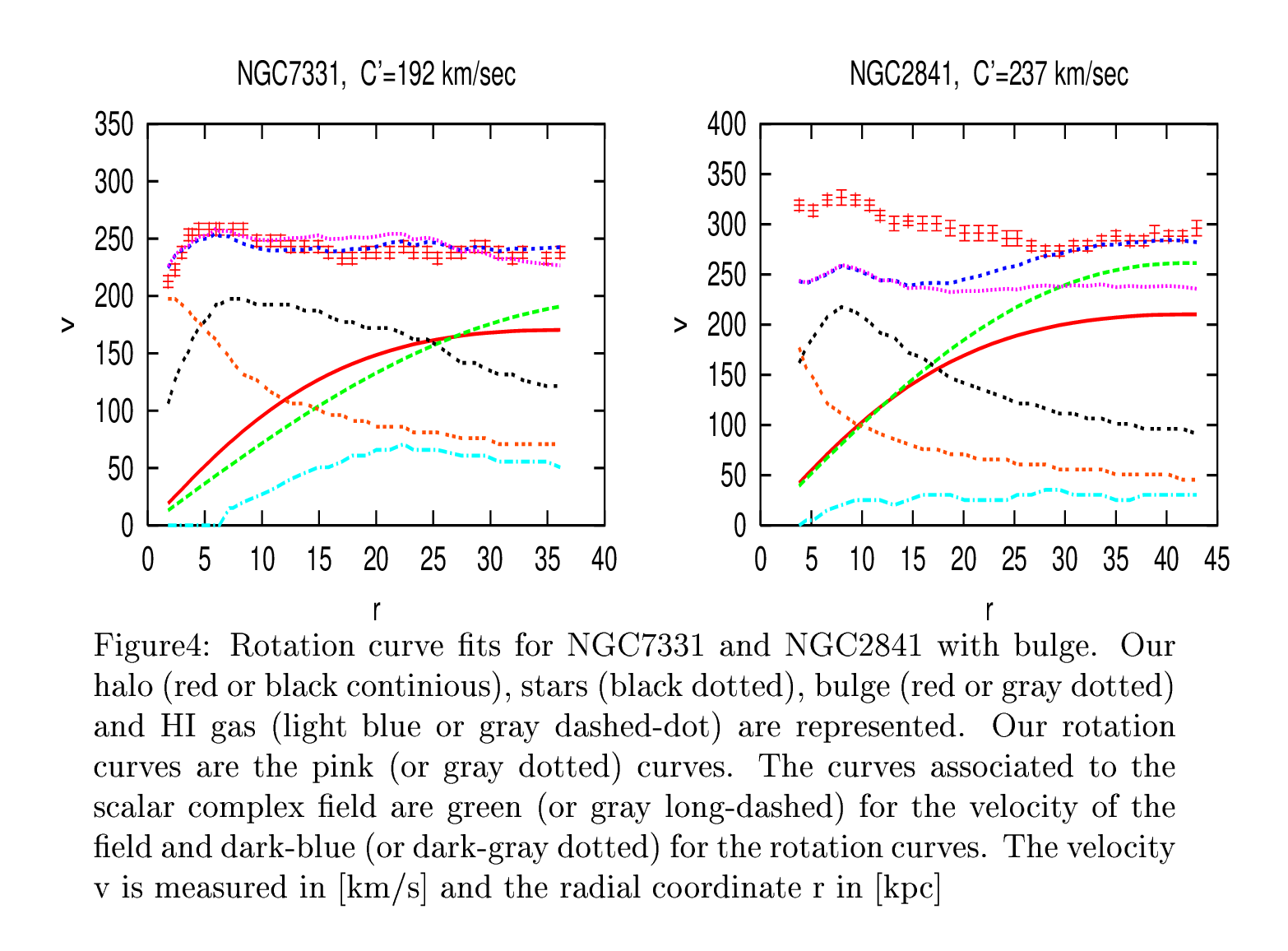}

\end{figure}
\end{center}

Our halo contribution depends on the parameter \( C' \) only. At each absolute
magnitude \( M_{I} \) corresponds one \( C' \) value.

The main results are shown in Fig.2. This figure shows eight observational rotation
curves with different absolute magnitudes \( M_{I} \) as well as the fitted
curves using equation (\ref{19}). The individual contributions from disk and
the scalar field are also shown. From the observational data, we see that low
luminosity spiral galaxy has a rather increasing rotation curve and a high luminosity
spiral galaxy a rather decreasing rotation curve. This means that a spiral galaxy
with low luminosity is stronger dominated by a dark matter halo than a spiral
one with high luminosity. Our model has the same behavior. Indeed, for low luminosity,
the scalar field contribution is more significant than in the case of high luminosity.
This fact is related to the parameter \( C' \) which only determines the halo
far from the luminosity matter region. Unfortunately, for low luminosity corresponding
to \( M_{I}=-18.5 \) for the first and \( M_{I}=-20.9 \) for the last one,
we don't have a good agreement of the data. Our curves are below the data with
an order of magnitude of \( 20\% \) for the first and \( 15\% \) for the other
ones. For high luminosity we obtain a rather good agreement with the data.

\subsection{Spiral galaxies rotation curves}

In this subsection, we consider two type of individual spiral galaxies ( \cite{15}
and\cite{17}), some of them with bulge and the others one without. Four galaxies,
where the stellar disk, gas and halo components contribute to the rotation curves,
are shown in figure 3. In figure 4, we present two spiral galaxies in which
a clear bulge can be read off the light curve. For each individual galaxies,
in order to obtain the best fit for our rotational curves, the \( C' \) parameter
is deduced from the observational data and the disk component so that: \( C'\simeq v_{disk} \)
in the range where the observational rotation curve becomes flat. Moreover,
for each spiral galaxy, \( v_{\phi } \) reaches its maximum at the radial coordinate
\( r \) corresponding to the last observational point. The data for rotation
curve fits are listed in Table (\ref{tab1}).

\begin{table}
{\centering \begin{tabular}{|c|c|c|c|c|c|c|}
\hline 
\( Galaxy \)&
\( Type \)&
\( \begin{array}{c}
Distance\\
(Mpc)
\end{array} \)&
\( \begin{array}{c}
V_{max}\\
(km/sec)
\end{array} \)&
\( \begin{array}{c}
Luminosity\\
(10^{9}M_{\bigodot })
\end{array} \)&
\( \begin{array}{c}
R_{HI}\\
(kpc)
\end{array} \)&
\( \begin{array}{c}
C'\\
(km/sec)
\end{array} \)\\
\hline 
\( NGC2903 \)&
\( Sc(s)I-II \)&
\( 6.40 \)&
\( 216 \)&
\( 15.30 \)&
\( 24.18 \)&
\( 153 \)\\
\hline 
\( NGC3198 \)&
\( Sc(rs)I-I \)&
\( 9.36 \)&
\( 157 \)&
\( 9.00 \)&
\( 29.92 \)&
\( 116 \)\\
\hline 
\( NGC2403 \)&
\( Sc(s)III \)&
\( 3.25 \)&
\( 136 \)&
\( 7.90 \)&
\( 19.49 \)&
\( 109.5 \)\\
\hline 
\( NGC6503 \)&
\( Sc(s)II.8 \)&
\( 5.94 \)&
\( 121 \)&
\( 4.80 \)&
\( 22.22 \)&
\( 96 \)\\
\hline 
\( NGC7331 \)&
\( Sb(rs)I-I \)&
\( 14.90 \)&
\( 257 \)&
\( 54.00 \)&
\( 36.72 \)&
\( 192 \)\\
\hline 
\( NGC2841 \)&
\( Sb \)&
\( 9.46 \)&
\( 326 \)&
\( 20.50 \)&
\( 42.63 \)&
\( 237 \)\\
\hline 
\end{tabular}\par}

\caption{\label{tab1}Data for rotation curve fits}
\end{table}

Concerning the spiral galaxies without bulge, the rotation curves are presented
in figure 3. Figure 3 shows the observational data, the individual contributions
from luminous matter, the gas and our halo of four different spiral galaxies.
The comparison with a scalar complex field, \cite{17} (\( \Psi (r,\, t)=P(r)\exp (-i\omega t) \))
which plays only the role of dark matter, is also shown. The maximum observed
rotation velocity for the galaxy NGC6503 is \( v_{max}=121 km/sec \)
and the order of \( \sim 216km/sec \) for the NGC2903. In these curves, we
remark that the agreement between our model and the observational data, is sensitive
to the maximum observed rotation velocity. Indeed, for the spiral galaxy NGC2903,
we obtain a very good agreement of the data as for the scalar complex field.
When the velocity decreases, i.e. for NGC2403 (\( v_{max}=136km/sec \)) and
NGC3198 (\( v_{max}=157km/sec \)), our curves are below the data in the surrounding
region, with an order of magnitude of \( 10\% \) for the first one and \( 15\% \)
for NGC3198 whereas the complex field model describes the data perfectly. In
the case of the Modified Newtonian Dynamics (MOND) model \cite{19}, the rotation
curves agree well with the observed curve for NGC2403 and for NGC3198 the differences
are in the opposite sense (about \( 15\% \)). However, concerning the spiral
galaxy NGC6503, our model as well as the complex scalar field don't reproduce
the observational data.

The spiral galaxies, NGC7331 (\( v_{max}=257km/sec \)) and NGC2841 (\( v_{max}=326km/sec \)),
with bulge are presented in figure 4. For NGC7331, our model (like the scalar
complex field and the MOND model) reproduces the observational data perfectly.
Unfortunately, we obtain very bad results for NGC2841 in the inner regions (about
\( 30\% \)), as in the case of the scalar complex field. However, in the outer
regions the scalar complex field model gives good agreement with the data while
our model is lower (about \( 20\% \)). In the MOND \cite{19} model, the predicted
curve is significantly higher than observed in the inner regions (about \( 15\% \))
and comparably lower in the outer regions.

Finally, we conclude that the considered tachyonic field (eq.\ref{equ11}) \cite{14}
gives interesting results at galactic scales, for reasonable velocities. For
extreme velocities such as: \( 121km/s \) (NGC6503) and \( 326km/s \) (NGC2841),
our model as well as the complex scalar field (and MOND model) don't reproduce
the observational data. In this context, such field seems to be a viable candidate
to explain the observations at both small and large scales.

\section{Conclusion}

In this paper, we were interested in a tachyon field \cite{14} (arising from
string theory ref.\cite{5},\cite{12},\cite{13}) which can behave like dark
energy with negative pressure at large scales and can cluster gravitationally
on small scales. Such tachyon field seems to be a viable candidate to explain
the observations at both small and large scales. Here, we have studied the behavior
of this tachyonic field at galactic scales and more precisely its contribution
to the dark matter component of the universe. By simplicity, we have retained
only the Newtonian solution of the Lagrangian considered. In this context, we
have fitted the rotation curves of six spiral galaxies (two with bulge and four
without bulge). 

Our rotation curves show that this model gives interesting results. Indeed,
for the galaxies NGC2903 and NGC7331 we obtain a good agreement with the data
and reasonable results for the other ones. our results are comparable with the
curves obtained on the one hand from a complex scalar field, which only plays
the role of the dark matter at the galactic scales, and on the other hand from
the MOND model. To conclude, we think that such tachyonic field is a viable
candidate to explain the observations at both small and large scales.

\subparagraph{ACKNOWLEDGMENTS:}

This work was partly supported by Euro-Gdr SUSY. I am grateful to K.Jedamzik
and H.Reboul for several useful discussions. I am also thankful to P.Brax, M.Marcelin
and A.Riazuelo for informative comments.

\end{document}